\documentclass[runningheads]{llncs}
\usepackage{graphicx}
\usepackage{times}
\usepackage{url}
\usepackage{latexsym}
\usepackage{times}
\usepackage{todonotes}
\usepackage{listings}
\usepackage{hyperref}
\usepackage{subfig}
\usepackage{multirow}
\usepackage{comment}
\usepackage{url}
\usepackage{subfig}
\usepackage{amsmath} 

\usepackage{amssymb}
\usepackage{hhline}
\usepackage{tabularx}
\usepackage{enumitem}

\begin{document}
%
\title{Citation Recommendation for Research Papers via Knowledge Graphs}

%
%
\author{Arthur Brack\inst{1}\orcidID{0000-0002-1428-5348}
\and Anett Hoppe\inst{1}\orcidID{0000-0002-1452-9509}
\and Ralph Ewerth\inst{1,2}\orcidID{0000-0003-0918-6297}}

\authorrunning{A. Brack et al.}
%
\institute{TIB -- Leibniz Information Centre for Science and Technology, Hannover, Germany \and
L3S Research Center, Leibniz University, Hannover, Germany \\
\email{\{arthur.brack|anett.hoppe|ralph.ewerth\}@tib.eu}}

%
%
%
\maketitle              
\begin{abstract}
Citation recommendation for research papers is a valuable task that can help researchers improve the quality of their work by suggesting relevant related work. Current approaches for this task rely primarily on the text of the papers and the citation network. In this paper, we propose to exploit an additional source of information, namely research knowledge graphs (KG) that interlink research papers based on mentioned scientific concepts. Our experimental results demonstrate that the combination of information from research KGs with existing state-of-the-art approaches is beneficial. Experimental results are presented for the STM-KG (STM: Science, Technology, Medicine), which is an automatically populated knowledge graph based on the scientific concepts extracted from papers of ten domains. The proposed approach outperforms the state of the art with a mean average precision of 20.6\% (+0.8) for the top-50 retrieved results.

\end{abstract}

\section{Introduction}
\label{intro}
Citations are a core part of research articles as they enable the reader to position the novel contribution in the scientific context. Moreover, relating own contributions with relevant research via references can also improve visibility. 
In consequence, it is in the interest of authors to provide complete and high-quality citation links to existing research.
However, this task becomes ever more complicated since the number of published research articles has been growing exponentially in the recent years~\cite{bornmann15growth}.

Consequently, the recommendation of suitable references for a piece of scientific writing is an important task to (a) improve the quality of future publications, (b) help authors and reviewers to point out additional relevant related work, and (c) discover interesting links to other areas of research. 
Färber and Jatowt~\cite{Farber20} distinguish between (1)~\emph{local citation recommendation} which aims to provide citations for a short passage of text, and \emph{global citation recommendation} which uses the documents' full text or abstract as the input. Here, we  focus on the task of global citation recommendation.

Current best-performing approaches for global citation recommendation \cite{Bhagavatula18,Cohan20,Zhou20} leverage primarily the articles' text and the citation network as information sources. 
In this paper, we explore another source of information, that is the set of scientific concepts which are mentioned in the article. The assumptions are (1) that additionally to the article's text, these provide condensed evidence to the described problem statement, used methodology or evaluation metrics, and (2) that research papers which should be citing each other usually share a similar set of concepts.
Consequently, we investigate whether research KGs interconnecting research papers based on the mentioned scientific concepts are instrumental in improving citation recommendation. 
For this purpose, we propose an approach which combines automatically extracted scientific concepts from the research articles with existing approaches for citation recommendation. 
The approach is evaluated on a KG that has been automatically populated from papers of ten scientific domains~\cite{Brack2021Coref}.
Our experimental results demonstrate that our proposed approach consistently improves the state of the art with a MAP@50 (mean average precision of top-50 results) of 20.6\%  (+0.8).
To facilitate further research, we release all our corpora and source code: \url{https://github.com/arthurbra/citation-recommendation-kg}.

The remaining of the paper is organised as follows: Section~\ref{sec:related_work} reviews existing research KGs and approaches for citation recommendation. In Section~\ref{sec:citation_recommendation_approach} we describe our proposed approach. 
The experimental setup and results are reported in Section~\ref{sec:citation_recommendation_setup} and \ref{sec:citation_recommendation_results}, while Section \ref{sec:citation_recommendation_conclusions} concludes the paper and outlines future work. 

\section{Related Work}
\label{sec:related_work}

Here, we briefly review research KGs and approaches for citation recommendation.

\subsection{Research Knowledge Graphs}
\label{sec:rkgs}

Various KGs interlink research papers through metadata (e.g. authors, venues) and citations \cite{Farber2019TheMA,Lo2020}, or through research artefacts (e.g. datasets) \cite{Amir2017ResearchGB,Manghi_Bardi_Atzori_Baglioni_Manola_Schirrwagen_Principe_2019}. 
Other initiatives organise scientific knowledge in a structured manner with community effort, such as Gene Ontology \cite{Carbon2019}, WikiData~\cite{vrandevcic2014wikidata} with encyclopaedic knowledge, or Papers With Code~\cite{PWC} and Open Research Knowledge Graph~\cite{Jaradeh2019OpenRK} for research contributions.

Furthermore, various KGs have been populated automatically from research articles. Computer Science Ontology (CSO) is a taxonomy for computer science research areas~\cite{Salatino2019TheCS}. Kannan et al.~\cite{Kannan20} create a multimodal KG for deep learning papers from text and images and the corresponding source code. 
The AI-KG has been generated from 333,000 research papers in the field of artificial intelligence (AI)~\cite{Dessi2020AIKG}. It contains five concept types (\textit{tasks, methods, metrics, materials, others}) linked by 27 relations types.
The COVID-19 KG \cite{Wise20} has been populated from the Covid-19 Open Research Dataset \cite{Wang20} and contains various biological concept entities.
Brack et al.~\cite{Brack2021Coref} generate a KG for ten science domains with the concept types \textit{material, method, process}, and \textit{data}.

\subsection{Citation Recommendation}

In the following, we outline recent approaches for global citation recommendation. For local recommendation, we refer to the survey of Färber and Jatowt \cite{Farber20}.

Bhagavatula et al. \cite{Bhagavatula18} propose a neural network-based document embedding model to retrieve candidate documents for a query document via similarity search~\cite{Johnson17} and a ranking model to rerank the top-$k$ candidates. 
The document embedding model is trained via a triplet loss with the papers' abstract and title using a Siamese architecture. 
It learns a high cosine similarity between document embeddings of papers citing each other. 
The reranker estimates the probability that a query document should cite a candidate document using the abstract, title, and optional metadata (e.g. author, venue) as features. 
Cohan et al. \cite{Cohan20} propose a document embedding model named SPECTER (Scientific Paper Embeddings using Citationinformed TransformERs). The SPECTER model is trained with an approach similar to Bhagavatula et al. \cite{Bhagavatula18}. 
However, they use a BERT encoder~\cite{Devlin2018BERTPO} pre-initialised with SciBERT embeddings~\cite{Beltagy2019SciBERTPC}.
Furthermore, Cohan et al. omit the reranking step and obtain the ranked results directly via the document embeddings' cosine similarity. 

Graph-based representation learning approaches learn document embeddings via graph convolution networks on the citation graph \cite{Hamilton17,Kipf17,Wu19}. 
However, they require the citation network also at inference time.
Other approaches \cite{Caciularu21,Jiang19,Zhou20} frame citation recommendation as a binary classification task: given a query and a candidate paper, the model learns to predict whether the query paper should cite the candidate paper. The models learn rich relationships between the contents of the two documents via various cross-document attention mechanisms. However, in contrast to the document embedding models \cite{Bhagavatula18,Cohan20}, such binary classification models can not be used for retrieval but only for reranking the top $k$ results, since a query paper has to be compared with all other documents~\cite{Chang20}.

To the best of our knowledge, approaches for citation recommendation that exploit knowledge graphs with scientific concepts have not been proposed yet.

\section{Citation Recommendation via a Research Knowledge Graph}
\label{sec:citation_recommendation_approach}

As the discussion of related work shows, 
citation recommendation approaches have not exploited research KGs yet. To leverage research knowledge graphs for citation recommendation, we propose an approach to combine document embeddings learned from textual content and the citation graph together with scientific concepts mentioned in the document.

Let $KG=(D,E,V)$ be a KG, $D$ the set of documents, $E$ the set of concepts, $V \subseteq D \times E$ the set of links between papers and concepts, and $E_d \subseteq E$ the set of concepts mentioned in paper $d \in D$. Let $one\_hot(e_i) \in \mathbb{R}^{|E|}$ be the one-hot vector for concept $e_i$ in which the i-th component equals $1$ and all remaining components are $0$.
Now, we construct the \emph{concept vector} $c_d \in \mathbb{R}^{|E|}$ for paper $d$ as follows:
\begin{equation}
\label{eq:concept_vector}
    c_d = \sum_{e_i \in E_d} one\_hot(e_i)
\end{equation}
Furthermore, let $s_d$ be a document embedding of paper $d$ obtained via an existing document embedding model (e.g. SPECTER~\cite{Cohan20}). 
The \emph{vector representation} $\vec{d}$ of paper $d$ is the concatenation of the concept vector $c_d$ and the document embedding $s_d$:
\begin{equation}
    \vec{d} = [c_d, s_d]
\end{equation}

For a query paper $q \in D$ the task is to retrieve the top $k$ results such that papers to be cited appear at the top of the list. We use cosine similarity for retrieval and ranking:
\begin{equation}
    rank (q, d) = \cos(\vec{q}, \vec{d}) = \frac{\vec{q}^{\,\intercal} \cdot \vec{d}}{||\vec{q}|| \cdot ||\vec{d}||}
\end{equation}

\section{Experimental Setup}
\label{sec:citation_recommendation_setup}

In this section, we describe the experimental setup, i.e. the used benchmark dataset, baseline approaches, and the evaluation procedure.

\begin{table}[tb]
\centering
\small
\caption{Characteristics of the STM-KG~\cite{Brack2021Coref} per domain in terms of number of abstracts, the number of citation links within the KG, and the number of scientific concepts in the cross-domain and in-domain KG. 
The number of concepts used across multiple domains are denoted as MIX. The domains are: Agriculture (Agr), Astronomy (Ast), Biology (Bio), Chemistry (Che), Computer Science (CS), Earth Science (ES), Engineering (Eng), Materials Science (MS), Mathematics (Mat), and Medicine (Med).}
\label{tab:stm_kg_stats}
\resizebox{\textwidth}{!}{
\begin{tabular}{l|rrrrrrrrrrr|r}
         & Agr    & Ast    & Bio    & CS    & Che   & ES    & Eng   & MS    & Mat  & Med    & MIX & Total     \\ \hline
\# abstracts      & 7,731 & 15,053 & 11,109 & 1,216 & 1,234 & 2,352 & 3,049 &  2,258 & 665 &  10,818  & - & 55,485 \\                  
\# citations      & 1,670 & 1,853 & 1,347 & 171 & 151 & 477 & 677 &  375 & 65 &  2,116  & - & 15,395 \\                  
\multicolumn{13}{c}{cross-domain KG}\\
KG concepts      & 138,342 & 173,027 & 177,043 & 20,474 & 21,298 & 62,674 & 55,494 & 39,211 & 9,275 & 227,690 & 70,044 & 994,572\\
\hline

\multicolumn{13}{c}{in-domain KG}\\
KG concepts & 180,135 &	197,605 &	229,201 &	30,736 &	32,191 &	81,584 &	78,417 &	55,358 &	14,567 &	278,686 &- &	1,178,480 \\
\hline

\end{tabular}
}
\end{table}

\paragraph{Benchmark Dataset:}

Existing benchmark datasets for research paper citation recommendation (e.g. \cite{Bhagavatula18,Cohan20,Lo2020}) do not provide a research KG that interlinks papers with scientific concepts. 
Therefore, we use the STM-KG~\cite{Brack2021Coref} as our benchmark dataset whose characteristics are depicted in Table~\ref{tab:stm_kg_stats}.
It has been populated from 55,485 abstracts in ten different scientific, technical, and medical domains and comes in two variants: (1) in-domain KG that shares scientific concepts only between papers of the same domain to avoid ambiguity of scientific terms (e.g. neural network in medicine vs. computer science), and (2) cross-domain KG that shares scientific concepts also between domains. 

The KG contains 15,395 citation links within the KG in total, of which 2,200 citation links are across papers from different domains. For evaluation, analogous to related work~\cite{Bhagavatula18,Cohan20}, we use only papers that cite at least four papers within the KG which results in 720 query documents and 4,069 citations links.
In contrast to Cohan et al.~\cite{Cohan20}, we pursue a realistic approach like Bhagavatula et al.~\cite{Bhagavatula18}, i.e. we retrieve top-$k$ documents from \emph{all} documents in the corpus instead of using predefined candidate sets of 30 documents (5 cited and 25 uncited papers) for each query document.

\begin{figure}[t]
    \center{\includegraphics[width=0.6\linewidth]
        {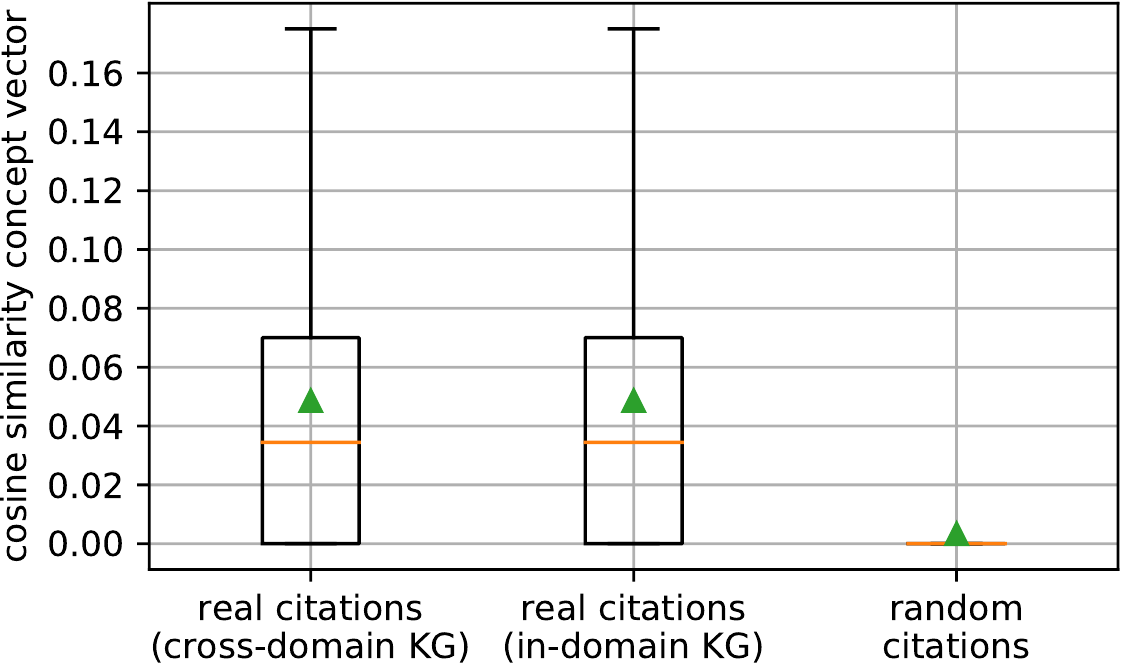}}
    \caption{Boxplot for cosine similarities between concept vectors of papers citing each other (15,395 links) for cross-domain and in-domain KG, respectively, and papers citing random papers (15,395 links). The green triangles depict the mean values.}
    \label{fig:boxplot_cosine_similarities}
\end{figure}

\paragraph{Baseline Approaches:}
We compare our approach with two simple (1 \& 2) and three strong baselines (3, 4 \& 5):

\begin{enumerate}
    \item \textbf{Random:} We use randomly initialised document embeddings with dimension 200.
    \item \textbf{Concept vector:} Only the concept vector is used for ranking (see Equation~\ref{eq:concept_vector}).
    \item \textbf{GloVe:} Document embedding of a paper is the average of GloVe~\cite{Pennington2014GloveGV} word embeddings obtained from the abstract of the paper. 
    \item \textbf{SciBERT:} Document embedding is also the average of the contextual word embeddings obtained from the abstract of the paper via SciBERT~\cite{Beltagy2019SciBERTPC} that is based on BERT~\cite{Devlin2018BERTPO} and has been pre-trained on scientific text. It has demonstrated superior performance in various downstream tasks on research papers~\cite{Beltagy2019SciBERTPC}.
    \item \textbf{SPECTER:} Document embedding is obtained via SPECTER~\cite{Cohan20} from the title and the abstract. The SPECTER model has been trained on the textual content and the citation graph of research papers, and is the current state of the art.
\end{enumerate}
To compute GloVe and SciBERT document embeddings, we use the \emph{sentence transformers} library~\cite{Reimers19}. For SPECTER we use the implementation of Cohan et al.~\cite{Cohan20}.

\paragraph{Evaluation:} To evaluate the quality of the ranking results for the top $k$ citation recommendations, we use \emph{Mean Average Precision} (MAP@k)~\cite{rankEval2021,MAP} as in related work~\cite{Cohan20}. MAP@k is the mean of the \emph{Average Precision at k} (AP@k) scores over the query documents. The metric AP@k assumes that a user is interested in finding many relevant documents and is thus an appropriate evaluation metric for citation recommendation:
\begin{equation}
    AP@k(q) = \frac{\sum_{k'=1}^{k} Precision@k'(q) \cdot rel(k')}{\mbox{\# relevant documents for }q}
\end{equation}
$Precision@k$ is the fraction of relevant documents among the top $k$ retrieved documents, and $rel(k)$ equals 1 if the document at position $k$ is relevant, 0 otherwise.

\begin{table}[t]
\centering
\caption{Experimental results (in percent) for citation recommendation with random vectors, only the concept vector as well as document embeddings obtained from GloVe, SciBERT and SPECTER with and without using the concept vector.}
\label{tab:citation_recommendation_results}
\begin{tabular}{l|rr|rr|rr}
                         & \multicolumn{2}{|c|}{MAP@10} & \multicolumn{2}{c|}{MAP@20} & \multicolumn{2}{c}{MAP@50} \\ \hline
Random                   & 0.0 &  &  0.0 &  & 0.0 &  \\ \hline
Concept vector (cross-domain KG)           & 7.5 &  &  8.0 & & 8.5 &  \\ \hline
Concept vector (in-domain KG)           & 8.1 &  &  8.7 & & 9.3 &  \\
- Material               & 3.7 &  &  4.1 & & 4.4 & \\
- Process                & 3.6 &  &  3.9 & & 4.2 &\\
- Data                   & 1.9 &  &  2.1 & & 2.2 &\\
- Method                 & 1.1 &  &  1.2 & & 1.4 &\\ \hline
GloVe                    & 9.1 &  &  10.0 & & 10.8 &   \\
GloVe + concept vector (cross-domain KG)   & 11.4 & (+2.3) & 12.5 & (+2.5)  & 13.4 & (+2.6)   \\ 
GloVe + concept vector (in-domain KG)   & 11.3 & (+2.2) & 12.5 & (+2.5)  & 13.5 & (+2.7)   \\  \hline
SciBERT                  & 10.2 & & 11.5 & & 12.6 &   \\
SciBERT + concept vector (cross-domain KG) & 12.1 & (+1.9) & 13.3 & (+1.8) & 14.4 & (+1.8)   \\ 
SciBERT + concept vector (in-domain KG) & 11.9 & (+1.7) & 13.2 & (+1.7) & 14.4 & (+1.8)   \\ \hline

SPECTER                  & 16.5 & & 18.3 & & 19.8 &   \\
SPECTER + concept vector (cross-domain KG)  & 16.9 & (+0.4) & 18.9 & (+0.6) & 20.5 & (+0.7)  \\
SPECTER + concept vector (in-domain KG)  & \textbf{17.0} & (+0.5) & \textbf{19.0} & (+0.7) & \textbf{20.6} & (+0.8)

\end{tabular}
\end{table}

\section{Results and Discussion}
\label{sec:citation_recommendation_results}

The boxplots in Figure~\ref{fig:boxplot_cosine_similarities} depict the distribution of cosine similarities of concept vectors between citing and non-citing papers. 
It can be seen that papers citing each other have on average a higher cosine similarity than papers not citing each other. 
This underlines our hypothesis that papers citing each other share a common set of scientific concepts.

Table~\ref{tab:citation_recommendation_results} shows the results of the evaluated approaches. Using only the concept vectors for ranking outperforms the random baseline significantly. When using only certain concept types (i.e. \textit{process, method, material}, or \textit{data}) we can observe that \textit{material} and \textit{process} concept types contribute most to the results. However, using all concept types together yields the best results.

Baseline ranking approaches via document embeddings learned from the text (GloVe and SciBERT), or text and the citation graph (SPECTER) outperform the ranking only via concept vectors significantly, while SPECTER performs best as expected. 
This indicates that concept vectors alone do not contain enough information for the task of citation recommendation. 
However, our proposed approach combining document embeddings and concept vectors consistently improves all baseline approaches. 
For SPECTER, the in-domain KG yields slightly better results than the cross-domain KG. However, in our error analysis we found out that concept vectors from the cross-domain KG provide more accurate rankings for cross-domain citations.

\emph{Our results indicate that the exploitation of a research KG as an additional source of information improves the task of citation recommendation.}

\section{Conclusions}
\label{sec:citation_recommendation_conclusions}
In this paper, we have investigated whether an automatically populated research KG 
can enhance the task of citation recommendation. For this purpose, we have combined document embeddings that have been learned from text and the citation graph together with concept vectors representing scientific concepts mentioned in a paper. The experimental results demonstrate that the concept vectors provide meaningful features for the task of citation recommendation. 
In future work, we plan to evaluate our approach on further research KGs and develop approaches that can learn document embeddings jointly from text, the citation graph, \emph{and} the research KG.

%
%
%
\bibliographystyle{splncs04}
\bibliography{references}
\end{document}